\documentclass[a4paper,12pt]{scrartcl}

\usepackage[utf8]{inputenc}
\usepackage[english]{babel}
\usepackage{amsmath,amsfonts,amssymb}
\usepackage[pdftex]{graphicx}
\usepackage{eurosym}
\usepackage{braket}
\usepackage[colorlinks=true, citecolor=black, linkcolor=black, urlcolor=black]{hyperref} 
\usepackage{authblk}
\usepackage{color}
\usepackage{multirow}
\usepackage{slashbox}
\usepackage{nameref}

\title{Dynamics of Human Cooperation in Economic Games}
\title{Dynamics of Human Cooperation in Economic Games}
\author[1]{Martin Spanknebel \thanks{martin.spanknebel@uni-bremen.de}}
\author[1,2]{Klaus Pawelzik \thanks{pawelzik@neuro.uni-bremen.de}}
\affil[1]{Institute for Theoretical Physics, University of Bremen}
\affil[2]{Center for Cognitive Sciences, University of Bremen}
\date{}
\begin{document}
\maketitle

\begin{abstract}
Human decision behaviour is quite diverse. In many games humans on average do not achieve maximal payoff and the behaviour of individual players remains inhomogeneous even after playing many rounds. For instance, in repeated prisoner dilemma games humans do not always optimize their mean reward and frequently exhibit broad distributions of cooperativity. The reasons for these failures of maximization are not known. 
Here we show that the dynamics resulting from the tendency to shift choice probabilities towards previously rewarding choices in closed loop interaction with the strategy of the opponent can not only explain systematic deviations from 'rationality', but also reproduce the diversity of choice behaviours. As a representative example we investigate the dynamics of choice probabilities in prisoner dilemma games with opponents using strategies with different degrees of extortion and generosity. We find that already a simple model for human learning can account for a surprisingly wide range of human decision behaviours. It reproduces suppression of cooperation against extortionists and increasing cooperation when playing with generous opponents, explains the broad distributions of individual choices in ensembles of players, and predicts the evolution of individual subjects' cooperation rates over the course of the games. We conclude that important aspects of human decision behaviours are rooted in  elementary learning mechanisms realised in the brain. 
\end{abstract}


\newpage
Humans and animals regularly fail to achieve maximal payoff even in simple choice situations and also when allowed for trying them out with many moves. Such observations have triggered a plethora of research into the lack of 'rationality' in human and animal decision behaviour. Deviations from utility maximization were explained by psychological peculiarities including e.g. probability matching \cite{Herrnstein1993} and mis-estimations of utilities \cite{Kahneman2003}. Also, auxiliary motives or emotions can have a substantial influence on decision behaviour which cannot always be excluded experimentally \cite{Henrich2001}. Last, not least exploration and learning mechanisms are increasingly suspected to contribute substaintially to deviations from strict rationality in repeated games (for an encompassing review see \cite{Erev2012}). Here, wide, skewed, and U-shaped distributions of choice propensities were occasionally observed \cite{Herrnstein1993,Erev2012,Chammah1965,Rapoport1966}.
 While these findings hint to the existence of multiple attractors for the choice propensities in analogy to similar considerations in evolutionary game theory \cite{Sigmund2011,Helbing2010}, this link has not been established yet for human learning in economic games.\\

 \begin{figure}  \centering
\includegraphics{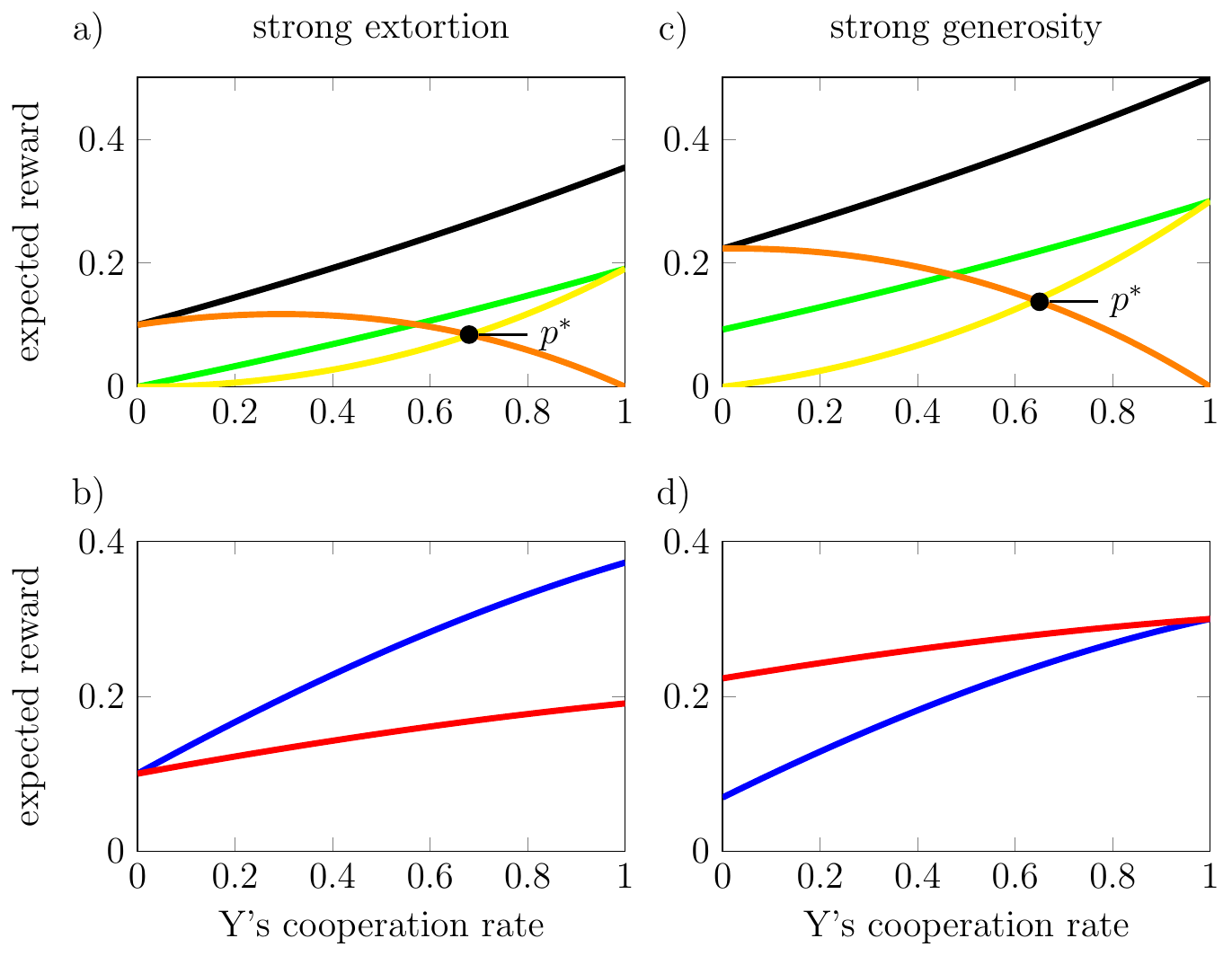}
\caption{Panel a) and c) show conditional expected rewards for strong extortion
and strong generosity, respectively, for the context free player using a single cooperation rate $p$. The black line is the expected reward r(p|y = C) given the player cooperated  and the green line the expected reward r(p|y = D) given the player defected. The yellow and orange lines are these conditional expected rewards weighted with their corresponding probabilities p and 1-p yielding the expected rewards from cooperative and defective acts, respectively. The learning rule has an unstable fixed point $p^*$ where these functions cross. Panel b) and d) show the total expected rewards of the player (red) and of the respective opponent ZD strategy (blue).}
\label{fig_explain_ZD}
\end{figure}

Social dilemma games have served as a paradigm in theoretical research of the evolution of cooperativity. Being particularly well understood they provide a testbed for investigating human deviations from optimal decision behaviour. In the prisoner's dilemma two players, named X and Y, are playing a game in which they can either cooperate or defect. In the case of mutual cooperation both get a reward $r_{CC}$. If X defects and Y cooperates, X gets an even larger reward $r_{DC}$ while Y gets an lower reward $r_{CD}$. If both players defect, both get an reward $r_{DD}$. These rewards must satisfy two equations: $r_{DC}>r_{CC}>r_{DD}>r_{CD}$ leads to a Nash Equilibrium of mutual defection, while $2r_{CC}>r_{DC}+r_{CD}$ guarantees, that mutual cooperation has the best global outcome. The latter condition establishes the dilemma between individual gain for defection and larger common benefit for cooperation. In the following we use the standard values defined in table \ref{table_reward}, which Axelrod used for his famous computer tournament\cite{Axelrod1984}.

\begin{table}[h]\centering
\caption{Definition of the reward r}
\begin{tabular}{cc|c|c} 
	 	                                                    & \multicolumn{1}{c}{} &  \multicolumn{2}{c}{\color{blue}{Computer: X}}                     \\ 
								    &		  	   &	cooperate $x=C$		&  defect $x=D$    	  \\ \cline{2-4}\parbox[t]{2mm}{\multirow{4}{*}{\rotatebox[origin=c]{90}{\color{red}{ \hspace*{1cm} Human:Y}}}}&  cooperate $y=C$	   & \backslashbox{\color{red}{0.3 ($r_{CC}$)}}{\color{blue}{0.3 ($r_{CC}$)}}	& \backslashbox{\color{red}{0.0 ($r_{CD}$)}}{\color{blue}{0.5 ($r_{DC}$)}} \\ \cline{2-4}
								    &   defect $y=D$	   & \backslashbox{\color{red}{0.5 ($r_{DC}$)}}{\color{blue}{0.0 ($r_{CD}$)}}	& \backslashbox{\color{red}{0.1 ($r_{DD}$)} }{\color{blue}{0.1 ($r_{DD}$)}} \\ \cline{2-4}    
\end{tabular}
\label{table_reward}
\end{table}

Recently a new class of strategies for the prisoners dilemma was introduced by Press and Dyson \cite{Press2012}, the so called Zero Determinant (ZD) Strategies. ZD strategies are memory-one strategies, i.e. they only use the information from the last round to base their decision. It can be shown that against ZD strategies longer memory strategies cannot have any advantage over memory-one strategies. ZD strategies have two further characteristics: (i) they enforce a linear relationship between the rewards of the players; (ii) for positive slope of this relation the best strategy for the co-player, meaning the strategy with the highest reward is total cooperation. In this paper we will concentrate on two extreme types of these ZD strategies: An extortion strategy which dominates any adaptive opponent \cite{Press2012}and a generous strategy which ensures that her own reward does not exceed the reward of the co-player \cite{Stewart2012}. Figure \ref{fig_explain_ZD} show the expected reward of a purely stochastic player (Y) having a single fixed cooperation probability playing when against the extortion \ref{fig_explain_ZD}b) or the generous strategy \ref{fig_explain_ZD}d) of X. For this simple strategy of Y, that takes no memory into account, but also for all other possible strategies of Y, total cooperation always leads to the maximum reward \cite{Press2012}.

An elegant experiment by Hilbe et\,al. demonstrated a particularly striking lack of cooperation in humans \cite{Hilbe2014} when repeatedly playing prisoner dilemma games against zero-determinant strategies that extort their opponents \cite{Press2012}. In contrast, when playing against generous ZD-strategies, human cooperation rates increased with time. As an explanation for this substantially sub-optimal behaviour against extortionists a desire for punishment was offered.\\
This experiment also revealed the large diversity of individual choice propensities ('cooperativities') sometimes leading to bimodal ('U-shaped') distributions after a series of moves, where some participants ended up always defecting while others always cooperated \cite[supplement, Fig. 1]{Hilbe2014}.\\
Obviously, the dynamics of choice propensities during the course of a game can lead to opposite behaviours in different individuals. We wondered if the coevolution of learning in the human player in interaction with the opponent's strategy could yield a dynamics that would explain this wide spread of individual behaviours.\\

Adaptations of human and animal behaviour in repeated games have been studied extensively in economics \cite{Kraines1993,Roth1998,Fudenberg1998}. Since the pioneering work of Pavlov more than hundred years ago research into reward dependent learning of choice preferences brought about a range of hypothetical learning rules based on success-dependent reinforcement of behaviour. It turned out that adaptations of animal as well as human behaviours are far from trivial even in very simple choice situations (for games see the review \cite{Erev2012}). Already the phenomenology of Classical Conditioning can be explained only partially by the famous Rescorla-Wagner learning rule \cite{Miller1995}. Therefore, more sophisticated algorithms for reinforcement learning were introduced, among which TD-learning \cite{Sutton1990} is now a leading paradigm in behavioural brain research and the Bush-Mosteller-Rule (BMR) \cite{Bush1953} is frequently used in Psychology. 

All psychological learning rules share the basic principle of reinforcing choice probabilities for successful choices and/or weakening choice probabilities for negatively rewarded decisions (Thorndike's famous 'law of effect',\cite{Thorndike1898}). Systematic biases either inherent in the particular learning rule or in specific parameter settings were frequently used to explain systematic deviations from optimal behaviour. As a representative example we here only cite a recent careful study in which success dependent learning effects in professional basketball players were explained by different learning rates for updating the respective choice probabilities \cite{Neiman2011}. \\

Several studies investigated adaptations of choice probabilities in their interaction with the dynamics of the environment. 
An important example here is the work of Herrnstein \cite{Herrnstein1993}, which shows that neither probability matching nor payoff maximization can explain human decision behaviour when the response of the environment is history dependent. To our knowledge the first mathematical analysis of the dynamics of choice probabilities was performed by Izquierdo and co-workers \cite{Izquierdo2008a}, who used the Bush-Mosteller-Rule to investigate the putative flow of cooperation probabilities of two players in social dilemma games. While their formal method is interesting from a theoretical perspective, the concrete predictions of this study are difficult to test since this would require knowledge of the strategies used by both human players which is not available. \\
The case considered by Hilbe et al. \cite{Hilbe2014} is more useful in this respect since the opponent's fixed zero determinant strategies are analytically fully transparent \cite{Press2012}. We use the data from these experiments as a benchmark for investigating the evolution of individual choice probabilities in interaction with a dynamic environment. \\

\section*{Results}

We model the choice behaviour $y(t) \in \{C,D\}$ of a subject Y in prisoner dilemma games using a stochastic strategy based on cooperation probabilities $p_Q(t)$. Given the context $Q$, and the decision at (discrete) time $t = 1, ..., T$ the player receives a reward $r(t)$ which depends on the behaviour $x(t) \in \{C,D\}$ of the opponent X. In general $Q$ can consist of any information available during the game. \\
Here we restrict our analysis to three cases:  $Q$ representing a) no context, b) the choice $x(t-1)$ of the opponent X in the last round, and c) the combination $(x(t-1), y(t-1))$ of the choices of X and Y in the last round. Since taking more history into account cannot help Y to increase her income when playing against Zero-Determinant (ZD) strategies  \cite{Press2012} it is not considered. \\

The choice probabilities $p_Q(t)$ are adapted according to the success of behaviour by a simple learning rule :

 \begin{align}
p_Q(t+1) = p_Q(t) + \left \{
\begin{array}{l l}
&  + \varepsilon_C  r(t) \delta_{ Q, Q'} \text{ for } y = C \\ 
&  - \varepsilon_D  r(t) \delta_{ Q, Q'} \text{ for } y = D
\end{array}
\right. 
\label{eq_lernen_1}
\end{align}

where $	\varepsilon_{C,D}$ are the learning rates in the case of cooperation and defection, respectively. $\varepsilon_C > \varepsilon_D$ leads to a bias $b$ towards cooperation. $\delta_{ Q, Q'}$ is the Kronecker-Delta which equals one only if $Q$  matches the context $Q'$ that was present at time $t$, and is zero otherwise. \\

With no context (case a) the index $Q$ can be dropped. The expected change $\Braket{\Delta p} = \Braket{p(t+1) - p(t)}$ of the cooperation probability $p$ can be calculated (see \nameref{chapter_methods}) and depends on the difference of the expected reward for cooperation $\bar{r}_C$ and that of defection $\bar{r}_D$.  Fig. \ref{eq_lernen_1}, a,c shows the results for the games against extortionist and generous opponents, respectively. The fixed points $p^*: \bar{r(p^*)}_C = \bar{r(p^*)}_D$  for both cases are unstable, players with initial $p$ above this point will tend to increase their cooperation probabilities and vice versa. In ensembles of players with widely sprad initial conditions the distribution will split, the respective cooperation probabilities will move towards the boundaries, and after many rounds one expects 'U-shaped' distributions of cooperation propensities. Ensembles with initial conditions above the fixed point will converge towards cooperation, which has indeed been observed \cite{Xu2015}. However, since the fixed points for extortion and generosity are quite similar this case is not able to explain the results from Hilbe et al. \cite{Hilbe2014}.

Case a) serves to demonstrates the basic effect: reward dependent adaptation of $p$ can induce multistability and not neccessarily leads to maximization 
For this case of no history dependency of the players cooperation propensity the difference between extortionist and generous opponents is small. Since the unstable fixed point are nearly at the same position($r(p^*) = 0.69$ in the extortion and $r(p^*)=0.65$ in the generous case) human subjects would act similar if playing against the different treatments. 

  \begin{figure}  \centering
\includegraphics{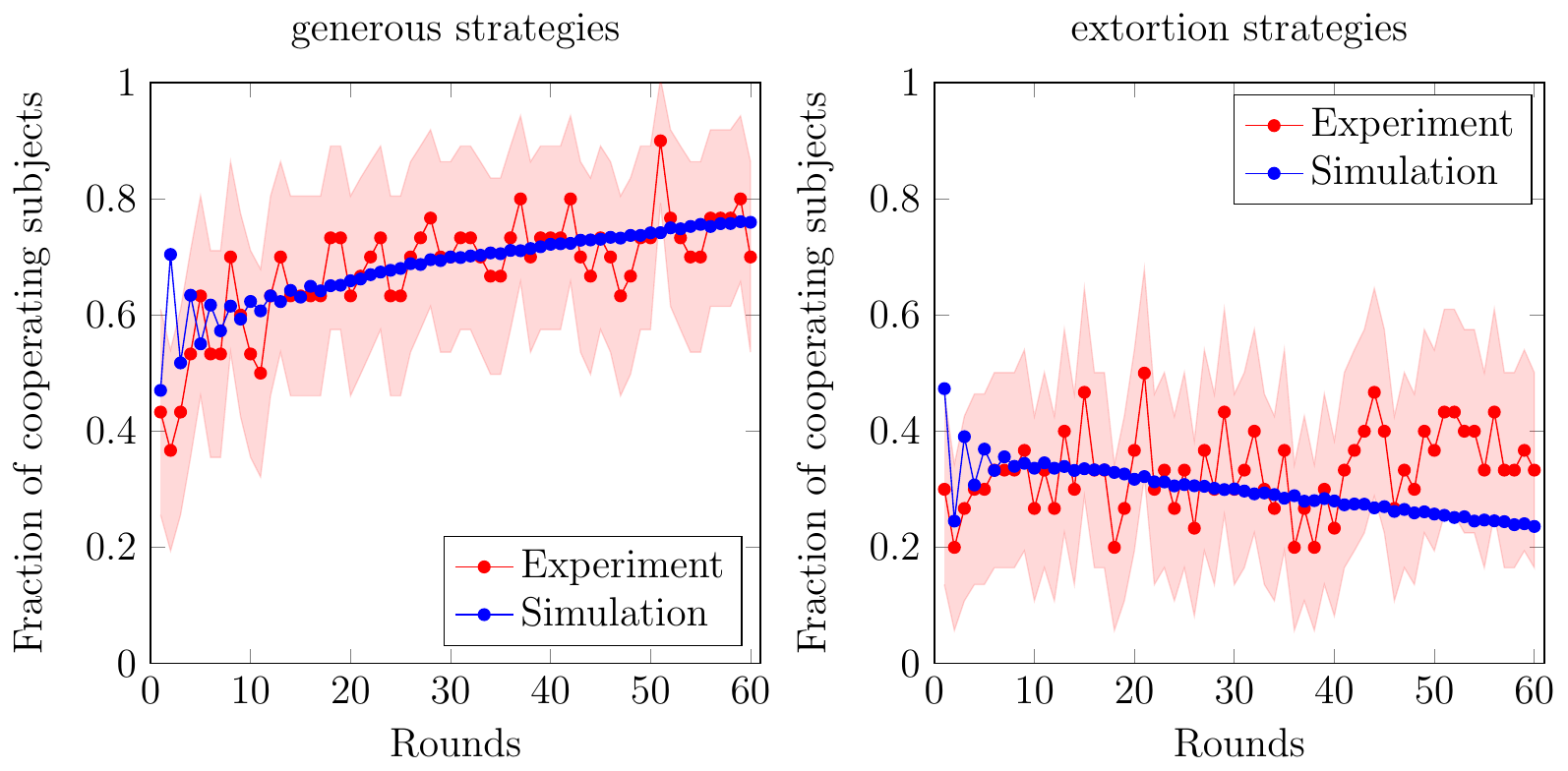}
\caption{Fraction of cooperating subjects during the course of the game from the experiments by Hilbe et\,al. \cite{Hilbe2014} (red) and in simulations (blue) with $N=10201$ players. The pink shaded region represents the standard errors from the experiments.}
\label{fig_cooperation_compare}
\end{figure}
Case b) with $Q \in \{C,D\}$ is more useful to explain the results by Hilbe et\,al. even if no cooperation bias is present $(\varepsilon_C = \varepsilon_D)$. Case c) leads to a similar dynamic as case b) and is therefore not shown. Figure \ref{fig_cooperation_compare} compares the fraction of cooperating subjects during the course of the game playing against extortion and generous treatments with the respective simulations of the model. In the experiment \cite{Hilbe2014} the humans start with a cooperation rate of 30\,-\,40\,\%. In the generous treatment the subjects behave ''rationally``, i.e. they optimise their strategy towards cooperation. At the end they reach a cooperation probability of 70\,-\,80\,\%, which leads to a mean payoff of \EUR{0.27} which is not far away from the best possible payoff \EUR{0.30}. Against extortion strategies humans behaved differently. Their cooperation probability dropped to low values (30\,-\,40\,\%) and they received only a mean reward of \EUR{0.15} which is quite far away from the maximally possible payoff \EUR{0.22}. The simulations lead to a very similar behaviour, in the generous case the cooperation probability starts at  47\,\% and goes up to 76\,\%. In the extortion case the agents also start with a cooperation probability of 47\,\% which decreases during the course of the game and ends with a probability of 24\,\%. In both cases the respective propensities of cooperation are fully consistent with the data, with an increasing cooperation probability in the generous and a slightly decreasing cooperation probability in the extortion case.

  \begin{figure} \centering
\includegraphics{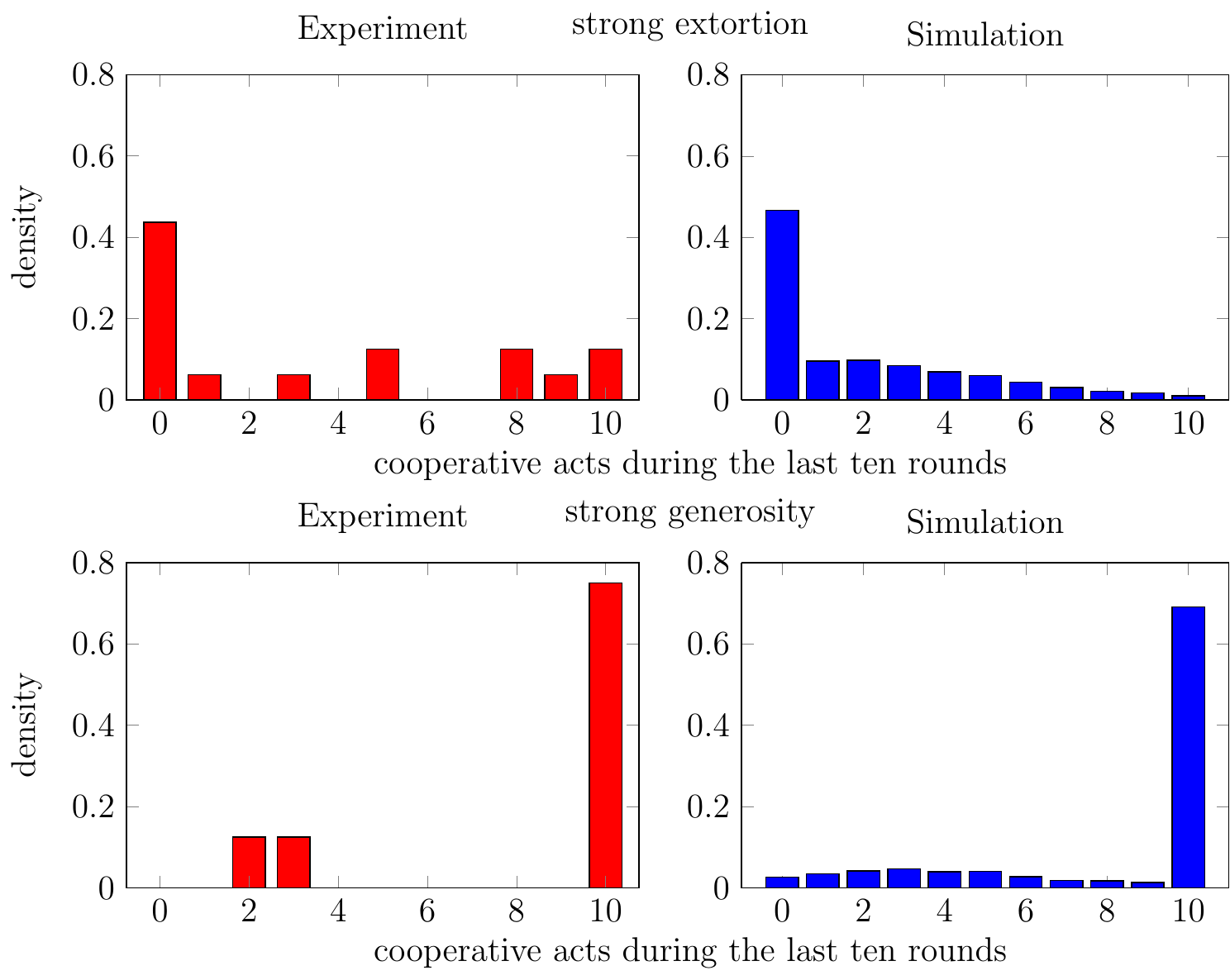}
\caption{Cooperation acts during the last ten rounds. Left part of the figure taken from \cite{Hilbe2014}.}
\label{fig_last_rounds}
\end{figure}

Figure \ref{fig_last_rounds} compares the behaviour of the model after 60 rounds with the results from Hilbe et\,al \cite{Hilbe2014}. In the strong extortion case the results of the model match the experimental data, the peak at zero has nearly the same height. The generous cased matches qualitatively, the large peak at ten appears in both, the simulation and the experiment, but in the simulation this peak is smaller than in the experiment.\\

\begin{figure}\centering
\includegraphics{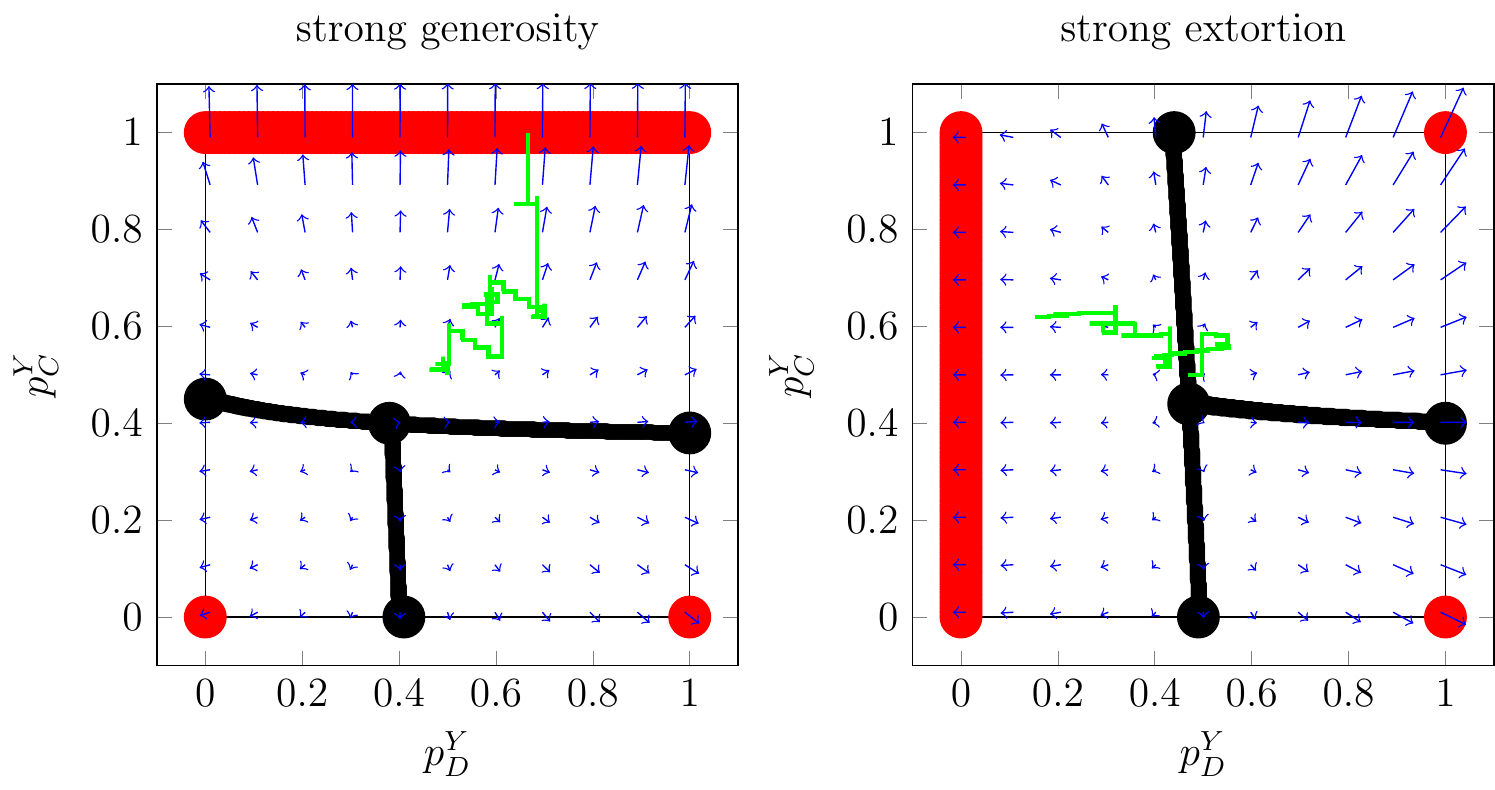}
\caption{Flow in the strong extortion and generous case with the unstable (black dots) and stable manifolds(red dots and lines) and the corresponding zones of attraction(black lines). The green curves are exemplary movements of single agents with the starting point $(0.5,0.5)$ over 200 rounds}
\label{fig_flow_fixpoint}
\end{figure}

To understand the dynamics of the model we calculate the expected changes of both choice probabilities $p_C$ and $p_D$ (see \nameref{chapter_methods}) and display them depending on $p_C$ and $p_D$ as a vector field (i.e. the flow in choice probability space as in Izquierdo et\,al. \cite{Izquierdo2008a}). Figure \ref{fig_flow_fixpoint} shows the unstable and stable fixed points and the corresponding basins of attraction of this flow. There are four unstable fixed points and three stable manifolds with basins of attraction that are delimited by the nullclines. 

For the strongly generous opponent we find a single unstable fixed point within the preference space at $(p_D^Y,p_C^Y)=(0.38,0.4)$. It determines an edge of the basin of attraction for the stable fixed point at the lower left edge at $(0,0)$, together with the fixed point at the left boundary at $(0,0.45)$ and at the lower boundary at $(0.41,0)$. This latter stable fixed point is the state of permanent defection. The basin of attraction for the stable fixed point at the lower right edge at $(1,0)$ is formed by the central and the left fixed point and the fixed point at the right boundary at $(1,0.38)$. At this fixed point the agents always take the different action as the opponent did in the last round, so we can call this behaviour ``anti-Tit-for-Tat''. 

The stable manifold most relevant for the experiment with generous opponents is the upper one at $p_C^Y = 1$. The corresponding basin of attraction is bounded by the left, central and right unstable fixed point. On the one hand the basin of attraction of this manifold has the largest overlap with the initial conditions (see \nameref{chapter_methods}), so most of the agents will go up towards this stable manifold. At the other hand agents in this manifold always cooperate if the opponent has cooperated in the last round, but because generous strategies also behave like this (see table \ref{table_coop_prob}) this is a point of permanent mutual cooperation. This explains why nearly every human in this experiment and the largest fraction of agents in the simulation cooperated permanently at the end of the game (see figure \ref{fig_last_rounds}).\\

\begin{table} \centering
\caption{Cooperation probabilities of the ZD strategies. $p_{CD}^X$ is the cooperation probability of the ZD strategy if the ZD strategy has cooperated and the opponent has defected in the previous round.}
\label{table_coop_prob}
 \begin{tabular}{|c|c|c|c|c|}
\hline
  Strategies	 	& $p_{CC}^X$ & $p_{CD}^X$& $p_{DC}^X$	& $p_{DD}^X$     \\ \hline
strong extortion & 0.692	     & 	0.000	 & 0.538	& 0.000	                  \\ \hline
strong generous	 & 1.000	     &  0.182 	 & 1.000	& 0.364	                  \\ \hline 
 \end{tabular}
\end{table}

For the strong extortion opponents the fixed points are similar to those of the strong generous case. However, the central unstable fixed point (here at $(0.47,0.44)$) together with the upper $(0.44,1)$, the lower $(0.49,0)$ and the right $(1,0.4)$ unstable fixed points now divide the space into three distinct basins of attraction. All agents, which are initialized in the upper right basin will tend to go to the stable fixed point at $(1,1)$ of permanent cooperation. In contrast, the agents which start in the lower right basin will move towards the anti-Tit-for-Tat fixed point at $(1,0)$.

The agents, who start at the left basin will tend to go to the large stable manifold at $p_D^Y=0$, which is most important for understanding the experimental results for this case. On the one hand the basin encompasses the initialization area and on the other hand the agents always defect if the opponent has defected in the previous round. Since extortion strategies do the same (see table \ref{table_coop_prob}) this leads to a state of permanent mutual defection. This explains why the largest fraction of subjects in the experiment and agents in the simulation prefer defection at the end of the game(see figure \ref{fig_last_rounds}).\\

\begin{figure}\centering
\includegraphics{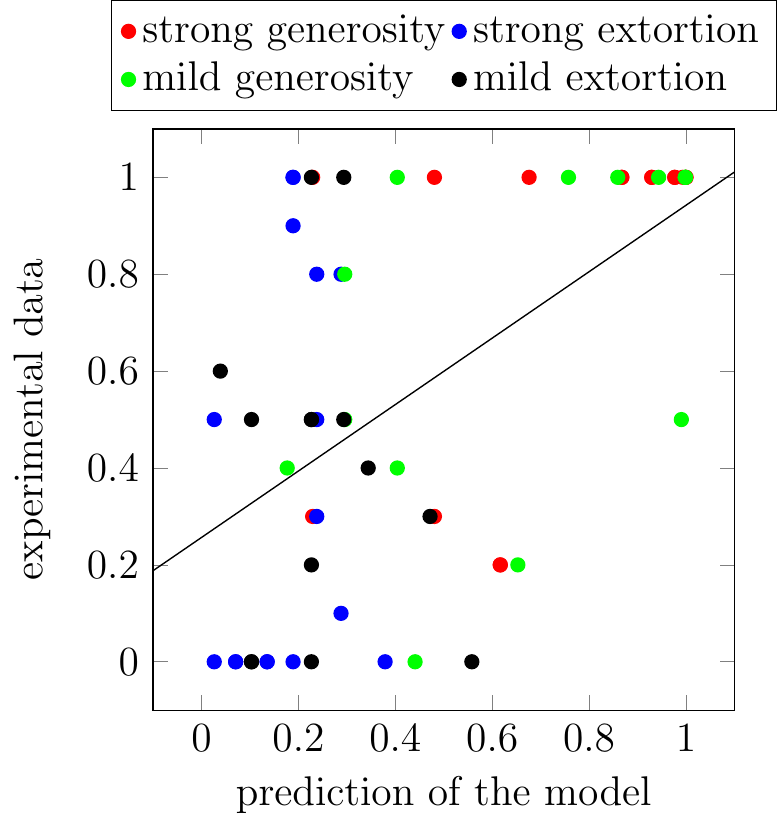}
\caption{Fraction of cooperative acts during the last ten rounds of each subject in all 4 ZD-treatments from the experiments of Hilbe et al. \cite{Hilbe2014} versus the predictions of the model. The solid line is the linear regression having a slope of 0.69 and a y-intercept of 0.26.}
\label{fig_density_human}
\end{figure}

A critical prediction of our model is that the evolution of the players choices should depend on their respective individual initial conditions. We tested this property of our model by comparing the final cooperation rates of individual subjects with the corresponding predictions of our model which were based on the subjects' observed choices during the first ten rounds (see \nameref{chapter_methods}). Figure \ref{fig_density_human} summarises the results for all games and every subject. The linear regression (solid line) leads to a function with a slope of $0.69$ and a y-intercept of $0.26$.\\

In principle, the flow field can be estimated directly from experimental data which will allow to experimentally test different hypothetical learning rules. To demonstrate this, we used a naive method (see \nameref{chapter_methods}) with computer generated data and found that a good match with the analytic flow can be achieved already from 500 agents in a game of 200 rounds (figure \ref{fig_measure_flow}).

\section*{Discussion}


The dynamics of choice probabilities was analysed for a simple model of learning in subjects playing iterated prisoner dilemma games against zero-determinant strategies. The flow in choice preference space exhibits characteristic basins of attraction: depending on their initial readiness to cooperate after many rounds of the game players tend to either always cooperate or to always defect. Thereby small individual differences in cooperation rates may lead to opposite behaviours in the long run. While the insight that learning can lead to extreme behaviours is not new \cite{Izquierdo2008a, Macy2002}, the general method introduced here allows to determine the potential influence of the dynamics on systematic deviations from optimal behaviour for all environments that depend on few past moves. For instance, we suspect that many of the systematic deviations from both, utility maximization and matching observed experimentally as in \cite{Herrnstein1993} can be explained by our framework.

When playing against opponents with zero-determinant strategies a longer memory beyond the last move cannot improve the payoff \cite{Press2012}, an intriguing fact, which makes these games a benchmark for investigating the joint dynamics of learning and environment. With a plausible initial distribution of cooperation probabilities the model is able to quantitatively explain the full phenomenology of the population data gathered by Hilbe et al. \cite{Hilbe2013}. In particular, it reproduces the average dynamics of cooperation rates and explains the wide distributions of cooperativity among individual players.

Our results are robust with respect to many details of the learning rule. As a test we applied the full Bush-Mosteller Rule \cite{Bush1953} and obtained similar results when the aspiration parameter in the BMR was small. The aspiration is an offset to which the received payoff is compared and only the difference is used for reinforcement of the choice probabilities. Since the subjects in the experiment of Hilbe et al. were payed according to the absolute payoff of the prisoners dilemma game and had no explicit costs setting this parameter to zero is a plausible choice. If, however, the subjects would need to compare the received rewards to some explicit costs our model would predict the emergence of stable fixed points in preference space located at intermediate values of cooperation rates. This prediction of our approach could easily be tested in future experiments.

A central prediction of our model that can be tested already on the data available from \cite{Hilbe2013} is that the asymptotic cooperation rates of individual players should depend on their initial tendency to cooperate. This critical prediction was confirmed to a striking degree: based on the sequence of choices of individual subjects during the first ten rounds of the games the model's predictions closely matched the average fraction of cooperative acts in the last ten rounds (figure \ref{fig_density_human}). Prediction improved with a non-zero bias for successful cooperation in the learning rule, which is equivalent to assuming an additional value of mutual cooperation \cite{Rabin1993}. These results demonstrate that the proposed learning mechanism not only qualitatively captures the dynamics of choices. We did not attempt to further optimize the two free parameters of the model (learning rate and cooperation bias), neither did we perform an extensive search for better initializations of the cooperation rates which obviously can only improve these results. Also, we did not consider different individual learning rates, which can further improve the match to experimental data, however, at the cost of additional model parameters. 

We believe, however, that the prediction performance of our model will improve more substantially if the initial cooperation propensity and the learning rate could be estimated directly for each subject before the game starts. Then, given the model is correct, only the noise intrinsic to players and opponents would limit its predictability. In principle, also the flow itself can be estimated from experimental data. Using a naive method flow estimation in simulations (see \nameref{chapter_methods}) became significant with 500 players making 200 moves. This second approach demonstrates that experiments are within reach, that would directly test hypothetical learning rules. We plan to perform these tests in the next future. Because the method of determining the flow from groups of players can be implemented for any learning rule it can in principle also be used for determining characteristics of learning implemented in subjects from selected groups. We speculate, that it would thereby become possible to investigate the relation of pathological behaviour (as e.g. in addiction) to systematic aberrations of learning.

Taken together, our results suggest a fundamental role of the joint dynamics of learning and environment for explaining experimentally observed deviations from optimal decision behaviour that has previously been overlooked. 

\section*{Methods}
\label{chapter_methods}

\subsection*{Model}

Since the learning rule \eqref{eq_lernen_1} can lead to probabilities smaller than zero and larger then one we always set the cooperation probability to zero and one, respectively, if these boundaries are passed. The learning rates are $\varepsilon_C = 0.09375$ and $\varepsilon_D = 0.03125$ and all simulations were performed with $N=101^2=10201$ agents. For the initial conditions we assume, that human beings rather cooperate if there opponent has cooperated as well respectively defect if the opponent has defected as well. That is we choose a uniform distribution with the borders $0<p_D^Y<0.45$ and $0.45<p_C^Y<1$. In the first round we choose one of the two cooperation probabilities randomly.

\subsection*{Mean Field Theory}

To understand the results of the model, we calculate the expectation value $F_k = <\Delta p_k^Y>$ of equation \eqref{eq_lernen_1}. $\sigma_{Y,t}$ only depends on Y's behaviour in the current round, the reward $r_{xy}$ depends on both players behaviour in the current round and the Kronecker delta $\delta_{x'k}$ depends on  X's behaviour in the previous round. The behaviour of X and Y in the current round are called $x$ and $y$, respectively, and in the previous round $x'$ and $y'$. $F_k$ only depends on $x$,$y$ and $x'$. To calculate this expectation value we need the joint probability $\pi(x,y,x')$:
\begin{align}
 F_k \sim& \sum_{\{x,y,x'\}\in \{C,D\}} \pi(x,y,x') \sigma _y r_{xy} \delta_{ x' k} . \label{eq_mean} 
\end{align}
To calculate $\pi (x,y,x')$ we sum over the joint probabilities which include the full information of the last and current round:
\begin{align}
 \pi (x,y,x') = \sum_{y' \in \{C,D\}} \pi(x,y,x',y') = \sum_{y' \in \{C,D\}} \pi(x,y|x',y')  S_{x'y'}. \label{eq_state_appear}
\end{align}
$S_{x'y'}$ is the probability of the occurrence of the state $(x',y')$. A repeated prisoner's dilemma can be seen as a Markov process. The only problem is, that Markov processes assume constant transition probabilities which is not the case in our model because of the learning rule. We now assume that this process is a quasistatic process, which allows us to use the steady state probabilities of the Markov process. These state probabilities can easily be calculated by solving the following eigenvalue equation:
\begin{align}
 \textbf{S}
 \begin{pmatrix}
p_{CC}^Xp_{C}^Y   		&         p_{CD}^X  p_{C}^Y &         p_{DC}^X  p_{D}^Y 	&        p_{DD}^X p_{D}^Y  \\
p_{CC}^X(1-p_{C}^Y)	&     p_{CD}^X(1- p_{C}^Y)	&    p_{DC}^X (1- p_{D}^Y)	&   p_{DD}^X(1- p_{D}^Y) \\
(1-p_{CC}^X)p_{C}^Y 	&    (1-p_{CD}^X) p_{C}^Y 	&    (1-p_{DC}^X) p_{D}^Y 	&     (1-p_{DD}^X) p_{D}^Y \\
(1-p_{CC}^X)(1-p_{C}^Y)& (1-p_{CD}^X)(1- p_{C}^Y)	& (1-p_{DC}^X)(1- p_{D}^Y) 	& (1-p_{DD}^X)(1- p_{D}^Y)
\end{pmatrix} = \textbf{S}
\end{align}
where $p_{CD}^X$ is the cooperation probability of the ZD strategy, if the ZD strategy has cooperated and the human opponent has defected in the previous round. Quasistatic processes in general are processes which have an infinitely long break after each parameter change in which the system can equilibrate so that there is a permanent equilibrium. In our case of a repeated prisoner dilemma this means that after every round with a parameter change  we assume an infinite number of rounds with no parameter change so that the system can equilibrate. 

The probability $\pi(x,y|x',y')$ is the probability that in the current round the state $(x,y)$ occurs, if in the previous round the state $(x',y')$ occurred. This probability can be calculated with the known cooperation probabilities of X and Y:
\begin{align}
 \pi(C,C|x',y') &=  p^X_{x'y'} p^Y_{x'} \\
 \pi(D,C|x',y') &= (1-p^X_{x'y'})p^Y_{x'}\\
 \pi(C,D|x',y') &= p^X_{x'y'}(1-p^Y_{x'})\\
 \pi(D,D|x',y') &= (1-p^X_{x'y'})(1-p^Y_{x'})
\end{align}

\subsection*{Prediction of Final Cooperativity}

For predicting final cooperation rates $c_\text{e}^i$ from initial cooperation rates $c_\text{s}^i$ we calculate numerically the probability $P(c_\text{s},c_\text{e})$of the model that a randomly chosen agent has the cooperativity $(c_\text{s},c_\text{e})$. With this probability we can calculate the conditional probability $P(c_\text{e}|c_\text{s})$
\begin{align}
P(c_\text{e}|c_\text{s}) = \dfrac{P(c_\text{s},c_\text{e})}{\int_{0}^{1}P(\tilde c_\text{s},c_\text{e}) \text{d} \, \tilde c_\text{s}}
\end{align}

and the conditional expectation value
\begin{align}
E(c_\text{e}|c_\text{s}) = \int_0^1  P(\tilde c_\text{e}|c_\text{s}) \tilde c_\text{e}  \text{d} \, \tilde c_\text{e}
\end{align}

Plotting the conditional expectation value of the predicted cooperation rates in the last ten rounds versus the experimental data $(E(c_\text{e}|c_\text{s}^i),c_\text{e}^i)$ should lead to a distribution similar to the angle bisector if the model is correct.

\subsection*{Measurement of the Flow}

\begin{figure}\centering
\includegraphics{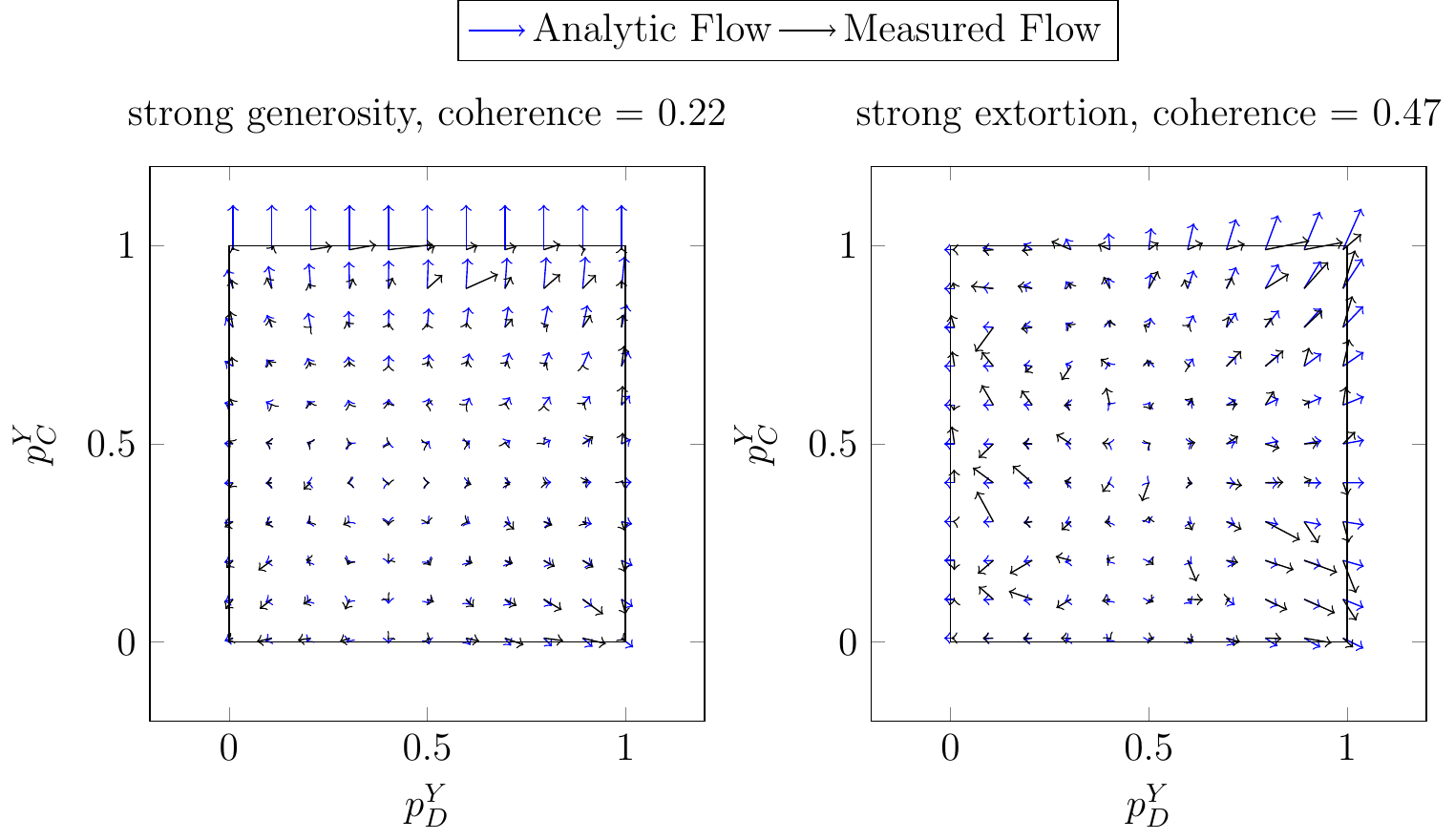}
\caption{Measurement of the flow for the strong extortion and strong generous strategies with computer generated date sequences of 500 agents and 200 rounds}
\label{fig_measure_flow}
\end{figure}

To check whether the learning rule is correct or not we want to estimate the flow from  experimental data. Therefore we divide the sequence into several parts (which can overlap) and calculate the mean value for $p_C^Y$ and $p_D^Y$ in each part.
\begin{align}
 \bar{p_C^Y} = \dfrac{n_{CC'}}{n_{CC'}+n_{DC'}} \\
 \bar{p_D^Y} = \dfrac{n_{CD'}}{n_{CD'}+n_{DD'}}
\end{align}
e.g. $n_{CD'}$ is the number of times the human subject cooperates in the current round, if the computer has defected in the previous round. If the sequences are long enough we can assume, that the differences of the mean values is caused by the change of the cooperation probabilities and not by noise. In the case that one agent only cooperates or defects it is only possible to estimate one of the two probabilities. In this case we assume that the other cooperation probability is unchanged and we use the probability of the previous part. If this appears at the beginning of the game we take the next probability.\\
We tested this very simple method with computer generated data of 500 agents in a game of 200 rounds. This is a large, but realistic experimental number of rounds. If the human subjects have e.\,g. 15 seconds for each decision, which is quite much for pressing one of two buttons, they would need 50 minutes. An experiment with 500 human subjects is a lot of work, but every subject plays independently against a computer. The humans do not have to play at the same time, so the amount of work increases linearly with the number of human subjects.\\
The results of this measurement are shown in figure \ref{fig_measure_flow}. To determine the quality of this method we calculate the coherence of the analytic flow $\textbf{F}_{\text a}$  and the measured flow $\textbf{F}_{\text n}$:
\begin{align}
c = \dfrac{\Braket{\textbf{F}_{\text a} \cdot \textbf{F}_{\text n}}^2}{\Braket{\textbf{F}_{\text a} \cdot \textbf{F}_{\text a}} \Braket{\textbf{F}_{\text n}\cdot \textbf{F}_{\text n}}}
 \end{align}
  The largest coherences were found at a part length of 110 rounds and an overlap between these parts of 90\,\% (so part one goes from round 1 to 110, part two from round 12 to 122 etc.). The coherence for the shown examples are 0.40 and 0.53, which are typical for the strategies. Especially at the border the analytic and measured flow do not match well. This is cause by the boundary condition which is not used in the calculation of the flow but in the computer generated data.

The coherences of the extortion cases are significantly higher than in the generous cases. This is caused by the left part of the flow ($p_D^Y = 0 \dots 0.5$) where nearly all arrows are parallel, so errors in $p_C^Y$ does not lead lo a lower coherence.
\section*{Acknoledgements}
We thank the autors of \cite{Hilbe2014} for sharing the data of their experiments, and Christian Hilbe for most fruitful comments on earlier versions of this manuscript. MS thanks the Friedrich-Ebert-Foundation for financial support.

\bibliography{Masterarbeit}
\bibliographystyle{pnas}
\end{document}